\documentstyle[preprint,epsf,epsfig,aps]{revtex}
\begin{document}

\author{A. Kabak\c c\i o\u{g}lu$^1$, I. Kanter$^2$, M. Vendruscolo$^3$, 
and E. Domany$^1$}
\address{$^1$Department of Physics of Complex Systems, 
Weizmann Institute of Science, Rehovot 76100, Israel \\
$^2$Department of Physics, Bar Ilan University, 52900 Ramat Gan,Israel \\
$^3$Oxford Centre for Molecular Sciences,
    Central Chemistry Laboratory \\
    University of Oxford,  
    South Parks Road, 
    OX1 3QH Oxford UK }
\title{Statistical properties of contact vectors}
\maketitle

\begin{abstract}
We study the statistical properties of {\it contact vectors}, 
a construct to characterize a protein's structure.
The contact vector of an $N$-residue protein is
a list of $N$ integers $n_i$, representing the number of residues 
in contact with residue $i$. 
We study analytically (at mean-field level) and numerically the amount of
structural information contained in a contact vector.
Analytical calculations reveal that a large variance in the contact numbers
reduces the degeneracy of the mapping between contact vectors and structures.
Exact enumeration for lengths up to $N=16$ 
on the three dimensional cubic lattice
indicates that the growth rate of number of contact vectors 
as a function of $N$ is only $3\%$ less than that for contact maps. 
In particular, for compact structures we present numerical evidence that, 
practically, each contact vector corresponds to only a handful of structures. 
We discuss how this information can be used for better structure prediction.
\end{abstract}

\begin{center} {\bf I. Introduction} \end{center}

The protein folding problem has been the subject of extensive
research in the last decade and although much has been learned
a satisfactory understanding of the phenomenon has not been
reached yet \cite{sali94,pande00,baker00}. 
The physical approach to the problem 
is to consider the native state of a protein as the
ground-state of a Hamiltonian which acts on sequence space and 
summarizes the inter-residue 
and residue-solvent interactions \cite{sali94,pande00,lazaridis00}.
Recently it was shown that there are several cases for which
there is no possible choice of pairwise contact 
interactions between residues
that suffices to pin down the native state 
even for a single protein~\cite{VD,VND}.
This conclusion is supported by molecular dynamics studies~\cite{md} 
and lattice models~\cite{Rios-Caldarelli} on 
residue-solvent interactions, where many-body forces are shown, 
or can be deduced to be, as relevant as two-body forces. 
To get around this failure of the two-body Hamiltonian approach
while retaining a coarse-grained description 
(as opposed to, say, an all-atom one, including water \cite{duan98a}),
we need to introduce new terms at the residue level, 
to bias the optimization procedure towards the true minima.

It is widely accepted that hydrophobicity is
the force driving the folding process \cite{dill95}.
At the individual residue level, hydrophobicity 
is correlated with the solvent-exposed surface area
in the native state \cite{Rost-Sander}. 
In addition, as reported below, a statistical analysis of the
native structures deposited in the Protein Data Bank (PDB) \cite{pdb})
reveals a good correlation (coefficient of correlation 0.8) 
between the solvent-hidden surface area per residue
and the number of inter-residue contacts per residue in the native state. 
We therefore propose the following two-step procedure 
for predicting the native state of a protein. 
First, a reasonably accurate prediction of the exposed surface area 
in the native fold is made on the basis of 
sequence information \cite{Rost-Sander}.
Second, this information is translated
into a prediction of the number of native contacts of each residue, 
e.g. to a predicted native {\it contact vector}.
Even if this scheme will turn out to be insufficient to perform
a successful prediction, it opens the possibility to
confine the search for the native fold to a small portion of the
conformational space. 
The question then becomes
``How many folded configurations are there, consistent with a
given set of contact numbers ?'' 
And, for that matter, ``Is such a contact-number
representation of the protein structure degenerate at all ?'' 
The rest of this paper addresses these questions.

\begin{center} {\bf II. Contact maps versus contact vectors} \end{center}
The contact map (CM) \cite{VSKDL} of a protein of $N$ amino-acids is a 
symmetric binary
matrix $C$ of size $N \times N$, such that $C_{ij}=1$ when the
$i^{th}$ and the $j^{th}$ amino-acids of the sequence are neighbors, 
with some suitable definition of ``neighborhoodness'' 
(e.g., a common construct is to threshold the pairwise
distance matrix for the C$_\alpha$ atoms \cite{VD}).
The CM has proven to be a convenient encoding of the 3-dimensional
native fold: 
1. The native backbone conformation can be reproduced to within
$\sim$ 1.5 {\AA} average uncertainty 
(the same as most X-ray data) \cite{VKD97}, 
2. It allows for an efficient search of the configuration space, 
since large conformational changes can be obtained by minor 
modifications of the CM \cite{VD98}.
Within such minimalistic framework 
one hopes to gain new insight to the protein-folding problem since it
is amenable to different physical and mathematical tools. 
For instance, the following Hamiltonian acting on the
contact map space has been extensively used in the
past~\cite{H-paper1,H-paper2,H-paper3,MJ-paper} :
\begin{equation}
\label{H}
H = \sum_{ij} { w(a_i,a_j)\,C_{ij} } \ ,
\end{equation}
where $w(\alpha,\beta)$ is one of the 210 energy parameters representing
the contact energy between the amino-acid types $\alpha$ and $\beta$.
Unfortunately, this formulation has limited predictive power. 
For example, 
given a large enough set of sequences and decoys (obtained by threading)
from the PDB, no set of $w(\alpha,\beta)$ exists, 
for which $H$ has its ground-states at the native-folds \cite{VND}.
This is in accordance with the recent studies
on the nature of the hydrophobic interaction~\cite{md,Rios-Caldarelli}, 
whose conclusion is that many-body interactions 
are of the same order of magnitude as two-body interactions.

One possible way to improve the Hamiltonian in Eq. (\ref{H}) is to include an 
energy penalty for deviations from the native-contacts:
\begin{equation} \label{H2}
H = (1-\lambda) \sum_{ij} { w(a_i,a_j)\,C_{ij} } + 
\lambda \sum_i (n_i -n^{nat}_i)^2 \\
\end{equation}
where we define a ``contact vector'' (CV)
$\vec{n}$ of rank $N$, which is the sum of the entries of 
the CM on each row (or column) (see Fig.(\ref{cm_cv_sketch})) :
\begin{equation}
\label{def_cn}
n_i = \sum_j C_{ij} \ .
\end{equation}

Contact vectors have already been studied in the context of protein folding
\cite{saitoh94,skorobogatiy97,ejtehadi98,li98c,hao98,pereira99,ejtehadi99,shahrezaei99,shahrezaei00}. We note in particular that
the second term in Eq. (\ref{H2}) resembles a hydrophobic 
term introduced previously~\cite{Mirny96} and studied in Ref. \cite{Park2000}, 
with the difference that there the desired number of contacts of residue $i$ is
determined by its species. 
Here instead we assume the knowledge of $n_i^{nat}$, the correct number of
contacts of residue $i$ in the native structure.
Hence the second term in Eq. (\ref{H2}) 
carries the same spirit as the Go model \cite{go}.
In this work we are interested in studying the statistical properties
of contact vectors. For our more general purpose,
it would seem inconsistent to use Eq. (\ref{H2}) 
to predict the native structure of a protein, as
we bias the Hamiltonian towards the minimum by using information which
is not accessible to us before we actually solve the problem. 
However, unlike in the Go model, 
the information required here about the native state (the
{\it number} of contacts for each residue) is modest, and, most crucially,
can be predicted.
Learning algorithms have been recently developed,
which are trained on known structures to predict the
surface exposure of the amino-acids 
in the native fold~\cite{Rost-Sander,fariselli01}.
Since the hydrophobic effect is driving the folding process \cite{dill95},
it is natural to expect that an accurate prediction of the solvent exposed
surface of each residue
in the folded state may lead to prediction of the correct native structure. 
To bridge the gap between the exposed-surface information and the CV defined
above, we performed an analysis on a representative set of proteins 
from the PDB database. 
We found a linear correlation with a coefficient of correlation
of 0.8 between the solvent {\it hidden} surface area of a residue
and the number of amino-acids it is 
in contact with (see Fig.(\ref{surf_vs_cont})). Therefore,
in future work
we expect to replace the $n_i^{nat}$ term in Eq. (\ref{H2}) 
by $n_i^{predicted}$,
thereby breaking the causality loop which is a characteristic
of Go-like models. 
Another reason to study the model of Eq. (\ref{H2}) is that
a related kind of Hamiltonian has been recently proved to be useful
to determine the structure of nearly-native
protein conformations \cite{vendruscolo01}. In that study,
$n_i$ represent the number of {\em native} contacts formed 
by residue $i$ in the contact map $C$. Also in that case, it was found
that a large variance in $n_i$ (see below) implies a low degeneracy in mapping
between contact vectors and three-dimensional conformations.

In studying Eq. ({\ref{H2}),
first we tried to use a set of contact energy parameters $w(a_i,a_j)$, 
found earlier by an optimization process, 
using Eq. (\ref{H2}) with $\lambda=0$.
This attempt failed to assign the  minimal energy to the native state 
for any choice of $\lambda$. 
However, an optimization of $w(\alpha,\beta)$ over the known
structures by using the Hamiltonian in Eq. (\ref{H2}) with $\lambda \neq
0$ may, perhaps, successfully identify the native state. 
We will investigate this possibility in the future.

The Hamiltonian (\ref{H2}), with 
$\lambda = 1$, fails to identify the native fold. 
This statement means that it is possible to find conformations which 
on the one hand are very different from the 
native one and, on the other, each amino-acid has 
exactly the correct number of neighbors, that is the same number of neighbors
as in the native state.
This result was first found by Ejtehadi {\em et al.} \cite{ejtehadi98}
by exact enumeration of all the compact conformations on a $3\times 3 \times 3$
cubic lattice. For actual proteins, 
an example is given in Fig. \ref{cm} 
in the case of protein CI2 (PDB code 2ci2), 
where the CMs of the native fold and of another conformation are superimposed. 
These two conformations have identical CVs.
At first glance, it would seem unlikely to find two compact 
configurations where each residue has
exactly the same number of neighboring residues in contact.
On the other hand, the cautious reader will attribute this degeneracy to
the loss of information (from $N(N-1)/2$ binary variables 
to $N$ integers of size $\leq N$) associated with going from 
a given CM to its corresponding CV via Eq. (\ref{def_cn}).
Quantifying the resulting degeneracy is a non-trivial problem.
The next section is an analytical attempt in this direction. 

\begin{center} {\bf III. An analytical approach} \end{center}

We ask the following question: ``How many contact maps exist for a 
given contact vector $\vec{n}$ ?''. In fact, we should be counting, for a given
$\vec{n}$, only the {\it physical} CM that are consistent with it. A physical
CM~\cite{Mirny96,VKD97} is one for which a perfectly matching chain 
configuration can be found. 
There is, however, no known analytical selection rule
for the physical CMs among all symmetric and traceless $N \times N$ matrices;  
therefore in our analytic study we will consider all binary symmetric matrices. 
This is essentially
the mean-field treatment of the problem, since in the limit of infinite
dimensions, all the constraints on the CM, except being symmetric
with zero trace, will be relaxed. For any finite dimension we overestimate
the degeneracy -  the number of physical CMs scales exponentially, 
as $e^N$, whereas the number of possible CMs scales as $e^{N^2} $ \cite{VSKDL}.

The formal expression for the number of symmetric, traceless
binary matrices consistent with a given vector, $\vec{n}$, is
\begin{equation}
d(\vec{n}) = \sum_{\{x_{ij}\}}^{i>j}\ \prod_{i=1}^N 
\delta_{(\sum_j x_{ij}),n_i}\ \ .
\end{equation}
The sum over $x_{ij}=0,1$ represents a
trace over all binary matrices, and the constraint $i>j$ ensures symmetry
and zero trace.
In order to perform the summation, we rewrite the Kronecker $\delta$ 
as a discrete
Fourier sum:
\begin{eqnarray} \nonumber
d(\vec{n}) &=& \sum_{\{c_{ij}\}}^{i>j}\ \prod_{i=1}^N \left[ {\frac{1}{N} 
\sum_{k=0}^{N-1}{ e^{i2\pi \frac{k}{N}(\sum_j{x_{ij}}-n_i)} } }  \right] \\
&=& \frac{1}{N^N} \sum_{k_1=0}^{N-1}\, \sum_{k_2=0}^{N-1}\, \cdots
 \sum_{k_N=0}^{N-1} \, \left(
\sum_{\{x_{ij}\}}^{i>j}  
{e^{i\frac{2\pi}{N}\sum_i k_i(\sum_j x_{ij} - n_i)}} 
\right).
\end{eqnarray}
Scaling $k_i$ by $N$, approximate the sums by
integrals. Then, evaluate the trace over the matrix elements, paying special
attention to $x_{ij} = x_{ji}$ and $x_{ii}=0$:
\begin{eqnarray} \nonumber
d(\vec{n}) &=& \int_{0}^{1}  dk_1 dk_2 ... dk_N \, \left(
\sum_{\{x_{ij}\}}^{i>j}  
e^{i 2\pi \sum_i k_i(\sum_j x_{ij} - n_i)}
\right)\ \\
&=& 2^{N(N-1)/2} \int_{0}^{1} dk_1 dk_2 ... dk_N \, 
e^{-i 2\pi \sum_i {k_i \left[(N-1)/2 - n_i\right] } }\, 
\prod_{i>j} \cos \left[\pi (k_i+k_j)\right]  \ .  
\end{eqnarray}
The integral can now be evaluated around its saddle points, $k_i = 1/2$
and $k_i=0,1$, which contribute equally. After we set $k_i = 1/2 + q_i$
and assume $N$ is divisible by 4, we obtain 
\begin{eqnarray} \
&\simeq& 2^{N(N-1)/2} 2 
\int_{-1/2}^{1/2} dq_1 dq_2 ... dq_N \, 
e^{-i 2\pi \sum_i {q_i \left[(N-1)/2 - n_i\right] } - (\pi^2/2) \,
\left[ N \sum_{i} q_i^2  + (\sum_i q_i)^2 \right]  }\  .  
\end{eqnarray}
The last square term in the exponent can be eliminated by a Hubbard-
transformation after rescaling $q_i$ by $\sqrt{N}$ and defining $\eta_i = n_i-(N-1)/2$:
\begin{eqnarray}  \nonumber
d(\vec{n}) &\simeq& 2^{N(N-1)/2} \sqrt{\frac{2}{\pi}}
\int_{-1/2}^{1/2} dq_1 dq_2 ... dq_N \, 
\int_{-\infty}^{\infty} dy e^{ - y^2/2 \, + \, i\pi y \sum_i qi 
\, + \, i 2\pi \sum_i { \eta_i q_i } \, - \, N\pi^2/2 \sum_{i} q_i^2 } \ ,  \nonumber 
\end{eqnarray}
which finally simplifies to yield
\begin{eqnarray}
\label{MF}
d(\vec{n}) &\simeq& 
\frac{2^{N^2/2}}{(N/\pi)^{N/2}\sqrt{N}} e^{-2\sigma_{\eta}^2 - 
\bar{\eta}^2 } \ ,
\end{eqnarray}
where $\bar{\eta}$ and $\sigma_{\eta}$ are the average and the standard
deviation of $\eta_i = (n_i-N/2)$. This is a mean-field estimate of
how the degeneracy of a CV scales with respect to the 
statistical properties of the CV. The leading behavior is clearly
far from being realistic, since the degeneracy should scale at most as
$z^N$ for some $z<z_{CM}$ ($\ln(z_{CM})$ is $0.83$ in 2d~\cite{VSKDL} 
and 1.32 in 3d as calculated here). 
Eq. (\ref{MF}) further suggests that the
maximally degenerate CV with a fixed average number of
contacts has $\sigma_{\eta} = 0$, i.e., all the amino-acids have equal
number of contacts, whereas an unbiased sample of CVs will be
dominated by those vectors with a typical standard deviation of
$\sigma_{\eta} \simeq \sqrt{N}$. The mean-field message is that the degeneracy
is a decreasing function of $\sigma_{\eta}$, i.e., variation in contact
number is desirable for low degeneracy. In the next section, we argue that
this is true away from the saddle point as well.

\begin{center} {\bf IV. Finite connectivity : Graph counting}\end{center}
In the previous section, we allowed for the number of contacts to take
any value between $0$ and $N$. In reality, and also in 
lattice models, the number of contacts is of order unity. Therefore, it is
desirable to have an estimate of the degeneracy of such CVs.
Once again, we consider all traceless, symmetric, binary $N\times N$ matrices.
We first observe that every such matrix encodes a unique 
graph with N
vertices, a vertex pair being connected if the corresponding matrix element
is 1. Symmetry ensures that the graph is undirected. We can ensure 
chain connectivity (but not the graph being physical!) by freezing
connections  on the first off-diagonal;  if we
choose to relax these ``backbone connections'', the remaining 
graph need not be connected.

The degeneracy of a CV, can then be
approximated by the number of graphs with N vertices and given
connectivities. We imagine the vertices from 1 to N with corresponding
number of legs sticking out of each and  we ask in how many ways these
legs can be connected such that none will be left out (the total number
of legs is an even number). Eq. (\ref{FD1}) follows immediately if one
imagines connecting pairs of legs sequentially (the numerator) and
remembering that legs coming out of the same vertex are interchangeable
(denominator).

Let's assume we allow the entries of the CV
to be one of $0,1,..,n$, $n\ll N$, and the composition given by
$\{p_0,p_1,..,p_n\}$, $p_i N \equiv N_i$ being the number of amino acids with 
$i$ contacts.
The average number of contacts is $ \sum_i i p_i \equiv c$.
The corresponding number of graphs reads
\begin{equation}
\label{FD1}
d(N,\{p_i\}) = \frac{(cN-1)!!}{(0!)^{N_0} (1!)^{N_1}...(n!)^{N_n}}\ .
\end{equation}
(The only difference with the usual Feynman diagram counting is the missing
$N!$ in the denominator: our vertices are distinguishable since they
correspond to the amino-acids labelled by their sequence number.)

Note that this expression is an approximation to the number of
symmetric traceless CMs, since diagrams with small loops involving one
vertex, as well as with more than one line connecting the same two vertices
{\it are} counted in Eq. (\ref{FD1}), even though they 
do not correspond to any CMs. However,
corrections due to excluding such diagrams do not change the scaling with 
$N$. 
Applying  Stirling's formula to Eq. (\ref{FD1}),
\begin{equation}
\label{FD2}
d(N,\{p_i\}) \simeq \exp \left[ \frac{cN}{2} \ln N + 
N ( \frac{c}{2} \ln c - 1 - \sum {p_n \ln (n!)} ) \right] \ .
\end{equation}
The leading order is now $z_F^{N\ln{N}}$ with $z_F = e^{c/2}$.
Better estimations require taking into consideration the spatial
correlations in the contact numbers due to the underlying one-dimensional
chain. 
Our next task is to find the compositions with the minimum and 
maximum degeneracy. The leading order in Eq. (\ref{FD2}) depends only on the
total number of contacts, so it is sensible to confine the search into the
subspace of CVs with a fixed average connectivity. 
We then extremize the next order term with respect to $\{p_i\}$, 
subject to the constraints $\sum p_m = 1$ and $\sum mp_m = c$ to find 
which distribution of contacts allows for the better ``designability'' 
(i.e., less degeneracy).
Fig.(\ref{p_min_max}) shows the choice of $\{p_i\}$ with maximum/minimum
degeneracy obtained numerically, as a function of the average contact number, 
$c$ (maximum number of non-backbone contacts, $n$, 
is chosen to be 4 as for the cubic lattice).
As read from the graphs, the highly degenerate scenario is when the number of contacts
for each residue is minimally away from the average, and vice versa for the low
degeneracy.
Even though here we deal with low connectivity $n_i << N$, whereas in the
previous section we had $n_i \sim N$, the result obtained here is
the same as there - low degeneracy  goes in parallel with maximal variation in
contact number.

One application of this principle is an order of magnitude estimation
for the ``optimal'' length for a protein. Consider a necklace model of
the protein, each residue represented by a sphere of fixed radius,
and the necklace
itself folded into a large compact sphere, where compactness is
imposed as a necessary condition for stability. Then, maximal contact
number fluctuation is attained when the number of buried residues
equals the number of residues on the surface. From this purely geometric
construction, one can estimate an ``ideal'' chain size:
Let the radii of the individual residues be unity, and the radius of the
protein be $R$. Assuming hcp-like packing, each residue occupies a volume
of $v=4\sqrt{2} (\mbox{unit})^3$, and those on the surface cover,
roughly, $a=4 (\mbox{unit})^2$ of surface area. Then, if $N$ is the number of
residues, we have $ v N = 4 \pi R^3/3$ and $ a N/2 = 4 \pi R^2 $,
which yields $N \sim 450$.

\begin{center} {\bf V. Numerical results}\end{center}

To compare the analytical findings presented above with numerical
simulations, we performed several exact enumeration studies on the square
and the cubic lattices. Therefore, in this section, we deal with {\it
physical} CMs and contact vectors, i.e., those generated by
self-avoiding walks in two and three dimensions.  In each analysis, we
kept a record of the distinct CMs we encountered and the corresponding
CVs.  Our first observation is that, the number of distinct CVs for a
given size $N$ scales exponentially with $N$ :
\[
N_{cv} \sim e^{a_{cv} N}\ .
\]
For given $N$, the number of CVs,
CMs, and self-avoiding walks increase in the given order. 
Yet, it is interesting that the growth rate $a_{cv}=1.28\pm 0.01$ is
only about $3\%$ less than the corresponding rate $a_{cm}=1.32\pm 0.01$
for the CMs in three dimensions (see Fig.(\ref{cv_exponent}), and also
\cite{VSKDL}).
The discrepancy between the mean-field analytical calculations and
the exact enumeration results points to the fact that the finite
dimensionality and the correlations between contacts due to the
underlying one-dimensional chain (i.e. working with physical CMs and CVs)
are crucial. 

The almost identical growth
rates is in accordance with our next analysis on the compact configurations
on a $6\times 6$ square lattice (see \cite{Tang}): Considering all the Hamiltonian walks inside
a $6\times 6$ square, we identified the number of walks that correspond to
each CV and found that the number of CVs with
degeneracy $d$ drops more or less exponentially with $d$ 
(see Fig.(\ref{cn_deg})). In fact, more than $96\% $ of all the contact
vectors have degeneracy $d \le 6$, although it is possible to find a vector
with 69 Hamiltonian walks mapped on it (not shown in Fig.(\ref{cn_deg})). 
The degeneracy gets
even smaller in the case of a compact but less than perfect packing
, in our case when the $6\times 6$
square is mostly filled with a 32-residue chain: introduced
vacancies (especially when in the core) ``label'' some of the residues
with otherwise identical contact number. Rearrangement of the core,
where all the residues have identical number of contacts, is the
dominant mechanism of
degeneracy. Hence, it gets more difficult to find conformations with
the same CV, once this degeneracy is lifted by the vacancies.
In this case, for practically all the CVs we have $d \le 5$.
In three dimensions, for the system sizes within reach, practically
all the residues are on the surface. Therefore, 
we have not extended this analysis to such case.

\begin{center} {\bf VI. Conclusion}\end{center}

Existing and future prediction methods for the
accessible surface area of individual residues can be
adopted to predict the number of native contacts of each amino acid
of a given protein. This prediction can then be used for an efficient search 
of the native contact map (and the corresponding conformation) in a 
dramatically reduced configuration space.
The prerequisite of such a program is to be able to identify different
folds consistent with a given set of contact numbers for each residue.
We investigated at the mean-field level the partition of the configuration
space (or rather the contact map space) into degeneracy classes labelled by
the CVs. The average degeneracy predicted by the
analytical calculations disagrees with the numerical findings, indicating
that the finite dimensionality and the correlations induced by the 
underlying one-dimensional chain are crucial even for a qualitatively
satisfactory result. 
We did find, already at the mean-field level, that
the increasing the fluctuations in the native contact-numbers 
reduces the contact vectors' degeneracy. 
This finding is also supported by another analytical calculation, 
valid in a different regime, where the average contact number is $O(1)$. 

We further investigated by exact enumeration the degeneracy spectrum of
CVs for self-avoiding walks on the square and the cubic lattice.
We found that for compact self-avoiding walks the CM and the
CV representations carry nearly the same amount of information.
This is an encouraging result, for an accurate enough prediction of
solvent exposed surface areas in the native state may then be used to reduce
the search space sufficiently, so that within the limited set of remaining
candidate CMs a simple pairwise interaction potential
may suffice to single out the native fold of the protein.
In addition, we performed exact enumeration over all SAWs of 
$N \leq 16$ steps in three dimension,
and found that the number of CVs grows exponentially with the protein
length, with a prefactor only a few percent smaller than that for the
CMs. The slow exponential growth of the average degeneracy of
the CVs is largely overestimated by our mean-field calculations.
Further analytical and numerical research is certainly called for.
We also observed that for {\it compact} configurations, CV $\rightarrow$ CM
mapping is practically one-to-few. The Hamiltonian in Eq. (\ref{H}), therefore,
may still be promising if the pairwise interactions are optimized within the
context of a (even roughly) predicted CV.

A. K. acknowledges many useful discussions with G. Getz and A. Punnose
and is also grateful to the Bilkent University Physics Department for their
hospitality during his visit. This work was partially supported by grants from 
the US-Israel Binational Science Foundation (BSF) and the Minerva Foundation.
I. Kanter thanks the Einstein Center for Theoretical Physics for partial
support.

\newpage

\begin{figure}
\caption{ The contact map is a binary representation of the three-dimensional
structure of the folded protein. The contact vector is constructed by summing
up the rows of the contact map.}
\label{cm_cv_sketch}
\end{figure}

\begin{figure}
\caption{ 
Hidden surface area per residue (after appropriate scaling) 
vs number of contacts per residue. 
The histogram is obtained by averaging over 177 representative proteins 
with a threshold on the $C_\alpha$ distance between amino-acid pairs 
\protect \cite{VD}.
The number of occurences is gray-scale coded, increasing from 0 to 500.
The coefficient of correlation is 0.8.}
\label{surf_vs_cont}
\end{figure}

\begin{figure}
\caption{ The native contact map (red) of protein CI2 (PDB code 2ci2) 
and a non-native
map (blue) with the same contact vector are overlapped. For clarity,
the symmetric half of the non-native contact map is omitted. }
\label{cm}
\end{figure}

\begin{figure}
\caption{ Solution of the mean-field equations for the maximal
and minimal degeneracy of contact vectors with finite average
contact number. Maximum contact number is chosen to be 4 as for the cubic lattice.
The solution for each $p_i$ is drawn within the
corresponding horizontal band. The y-axis of each band is labeled
on the left and right alternatingly. For fixed average contact number, $c$,
lowest degeneracy is when the standard
deviation in the contact numbers is maximal, and vice versa. }
\label{p_min_max}
\end{figure}

\begin{figure}
\caption{ Scaling of number of contact maps and vectors with chain length,
obtained by exact enumeration on the three dimensional cubic lattice. 
The inset (also
a log-plot) shows clearly that the deviation between the two growth rates is
real.}
\label{cv_exponent}
\end{figure}

\begin{figure}
\caption{ Degeneracty of the contact vector on a $6\times 6$ square. 
The upper graph is
a log-plot of the number of distinct contact vectors with the
degeneracy given on the $x$-axis. The lower graph shows the fraction of
SAWs of size 32 and 36 inside the square, covered by 
the subset of corresponding contact vectors with degeneracy $\le$ x. }
\label{cn_deg}
\end{figure}

\newpage

\begin{figure}
\epsfig{figure=fold_cm_cv.eps,width=15cm,angle= 0}
\end{figure}
\vspace{3cm}
{\Huge Figure \ref{cm_cv_sketch}}

\begin{figure}
\epsfig{figure=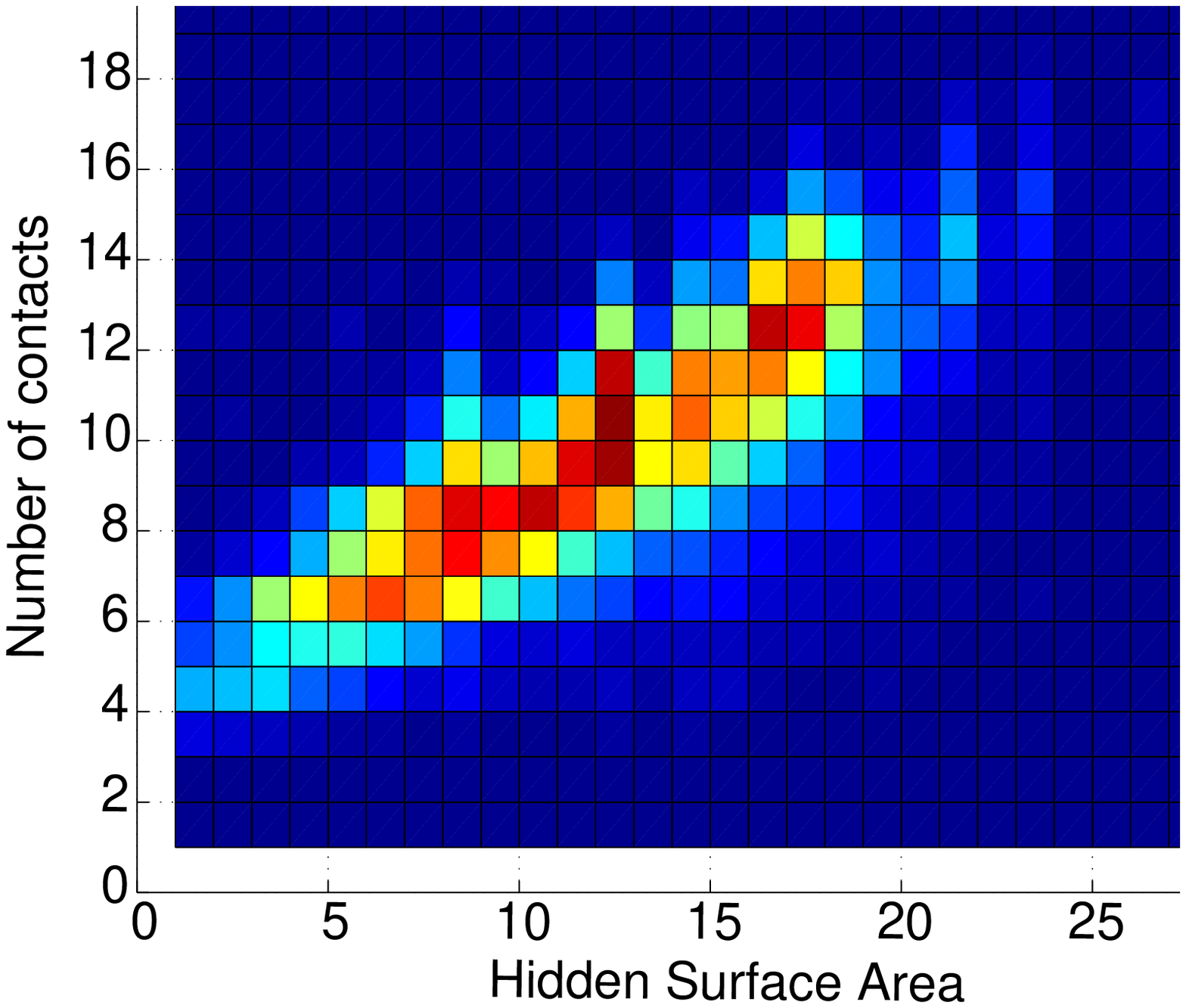,width=15cm,angle= 0}
\end{figure}
\vspace{3cm}
{\Huge Figure \ref{surf_vs_cont}}

\begin{figure}
\epsfig{figure=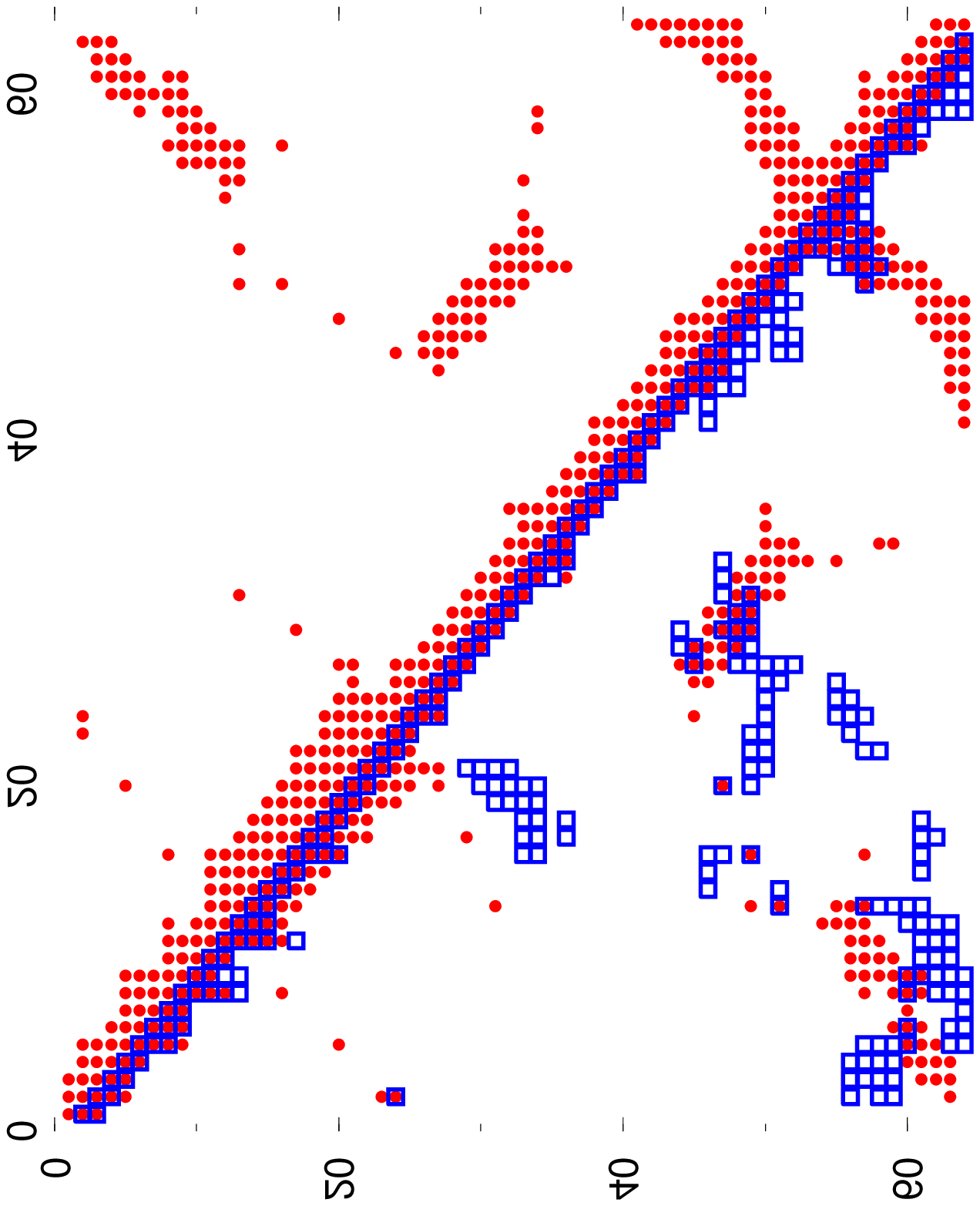,width=10cm,angle= -90}
\end{figure}
\vspace{1cm}
{\Huge Figure \ref{cm}}

\begin{figure}
\epsfig{figure=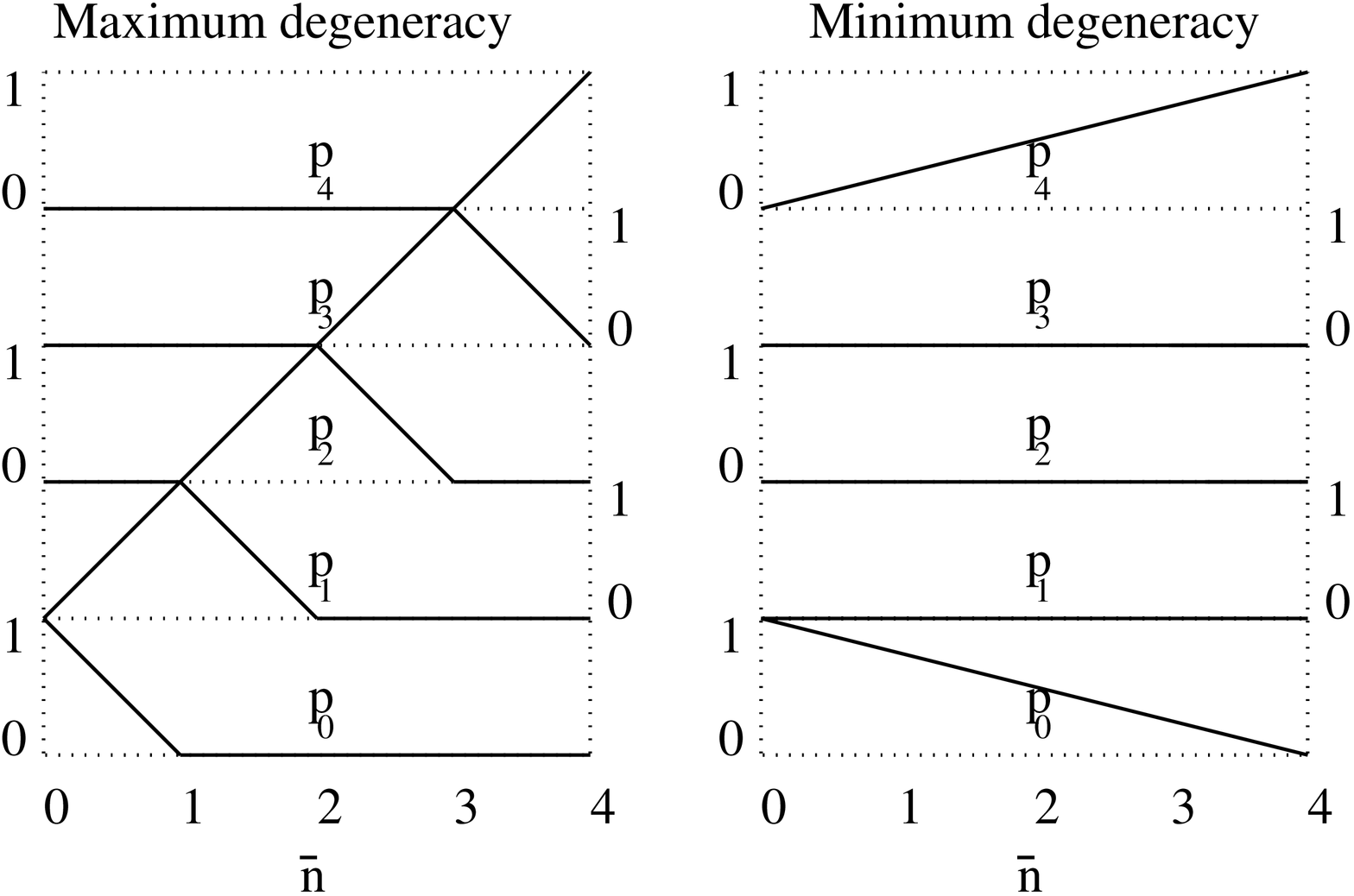,width=10cm,angle= -90}
\end{figure}
\vspace{3cm}
{\Huge Figure \ref{p_min_max}}

\begin{figure}
\epsfig{figure=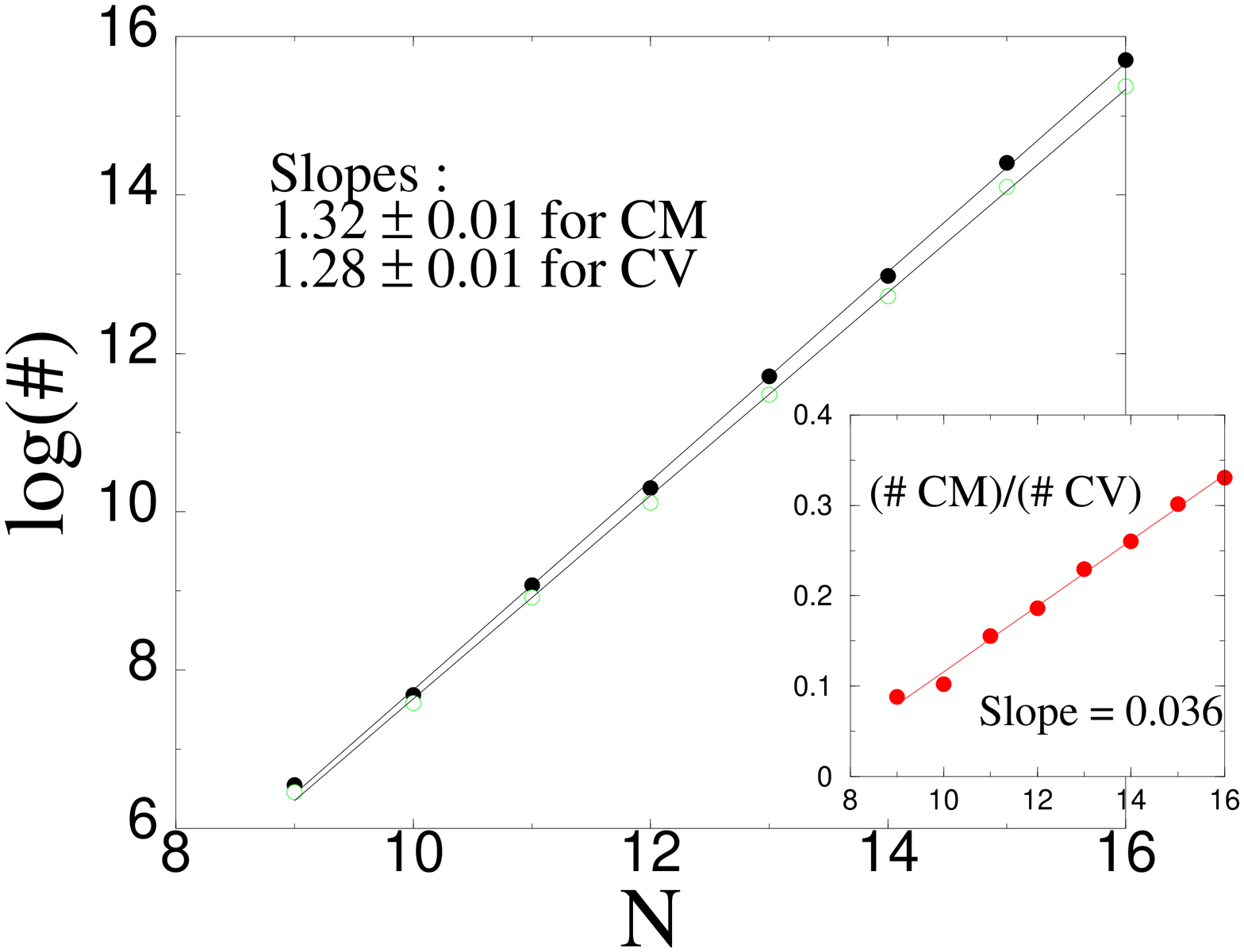,width=10cm,angle= -90}
\end{figure}
\vspace{3cm}
{\Huge Figure \ref{cv_exponent}}

\begin{figure}
\epsfig{figure=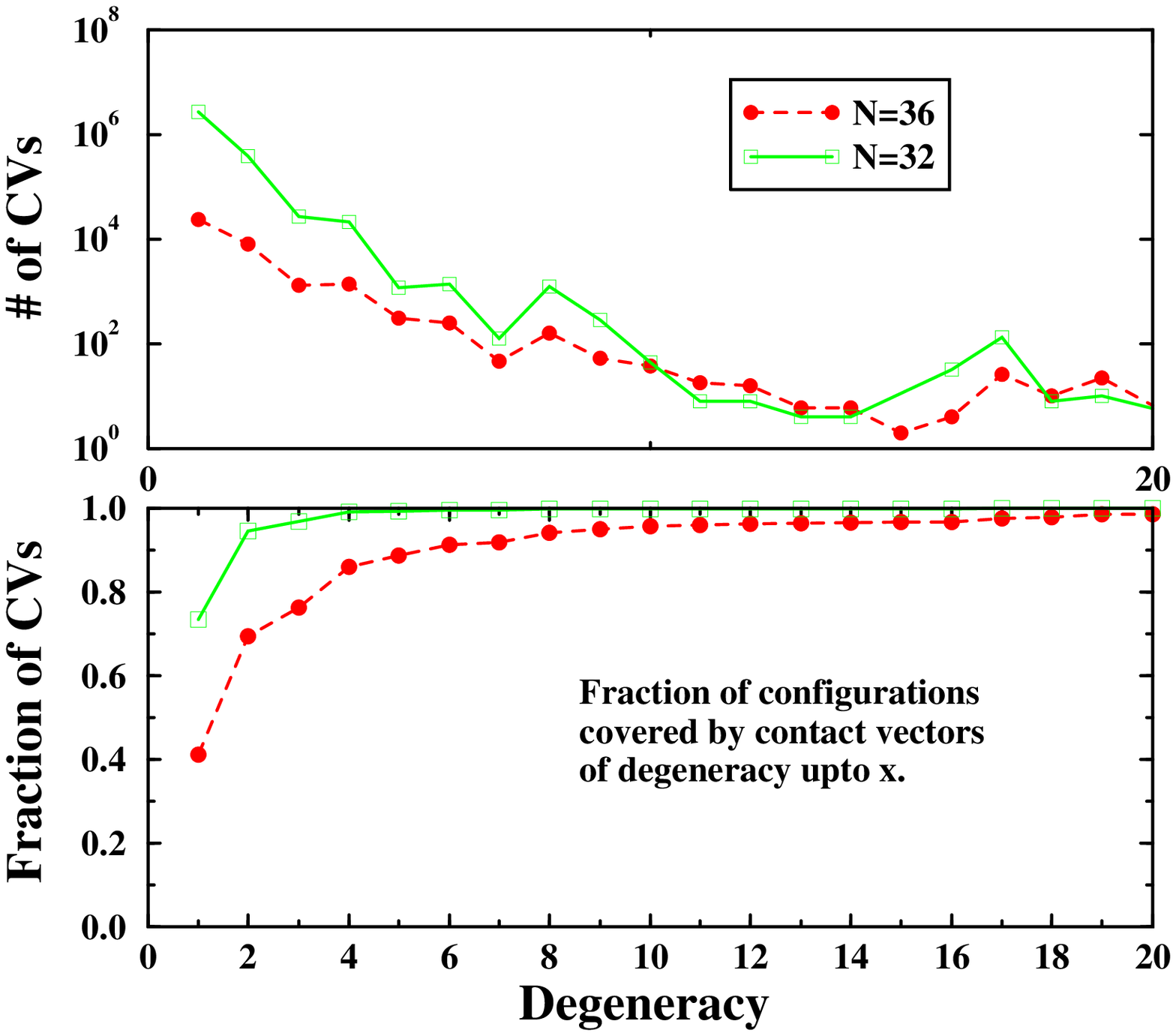,width=15cm,angle= 0}
\end{figure}
\vspace{3cm}
{\Huge Figure \ref{cn_deg}}

\end{document}